\documentclass[preprint,12pt]{elsarticle}

\usepackage{amsmath}
\usepackage{braket}

\usepackage{graphicx}
\usepackage{amssymb}
\usepackage{amsthm}
\usepackage{amssymb}
\setcitestyle{square}
\usepackage[toc,page]{appendix}
\usepackage{listings}
\usepackage{fancyvrb}
\usepackage{float}
\usepackage{xcolor}
\usepackage{subfig}
\usepackage[top=1.2in, bottom=1in, left=1.2in, right=1in]{geometry}
\usepackage{courier}
\usepackage{hyperref}
\usepackage{doi}
\usepackage{bbm}
\usepackage{bm}
\usepackage{booktabs}
\usepackage{siunitx}
\usepackage{mathrsfs}
\usepackage{braket}
\usepackage[space]{grffile}
\usepackage{tikz}
\usetikzlibrary{calc, intersections, decorations.markings}
\usepackage{ifthen}
\usepackage[normalem]{ulem}

\biboptions{comma,round}

\journal{ArXiv}

\begin{document}

\begin{frontmatter}




\title{Non-Equilibrium Phase Transition in a Spin-1 Dicke Model}
\author{Zhang Zhiqiang$^{1,2}$, Chern Hui Lee$^{1,2}$, Ravi Kumar$^{1,2}$, K. J. Arnold$^{1,2}$, Stuart J. Masson$^{3}$,\\ A. S. Parkins$^{3}$ and M. D. Barrett$^{1,2}$}
\address{$^1$Center for Quantum Technologies, 3 Science Drive 2, 117543, Singapore \\ $^2$Department of Physics, National University of Singapore, 2 Science Drive 3, 117551, Singapore \\ $^3$Dodd-Walls Centre for Photonic and Quantum Technologies, Department of Physics, University of Auckland, Private Bag 92019, Auckland, New Zealand}
\begin{abstract}
We realize a spin-1 Dicke model using magnetic sub-levels of the lowest $F=1$ hyperfine level of $^{87}$Rb atoms confined to a high finesse cavity. We study this system under conditions of imbalanced driving, which is predicted to have a rich phase diagram of nonequilibrium phases and phase transitions. We observe both super-radiant and oscillatory phases from the cavity output spectra as predicted by theory. Exploring the system over a wide range of parameters, we obtain the boundaries between the normal, super-radiant and the oscillatory phases, and compare with a theoretical model.
\end{abstract}


\end{frontmatter}


\section{Introduction}
The study of atom-light interactions has been an important area of research due to its importance in understanding of fundamental physics, especially quantum optics, and potential technological applications, such as quantum computation \cite{mabuchi2002cavity}, the quantum internet \cite{kimble2008quantum}, atomic clocks \cite{jiang2011making}, and superradiant lasers \cite{bohnet2012steady}. A significant amount of effort has been devoted to study coherent control of atoms by strong atom-photon interactions using various systems including optical cavities \cite{ye1999trapping}, atom chips \cite{colombe2007strong}, and structured optical waveguides \cite{goban2015superradiance}. An ensemble of cold atoms coupled to a cavity mode offers a suitable platform to study collective properties of atoms in a very precisely controlled environment \cite{arnold2011collective}. Such a system has been used to study spatial self-organizations of atoms \cite{arnold2012self}, superradiant lasing \cite{bohnet2012steady} and quantum phase transitions \cite{emary2003quantum, Baumann2010} based on the Dicke-model \cite{dicke1954coherence, garraway2011dicke}. Experiments on the simulation of the Dicke-model have mainly focused on spin-$\frac{1}{2}$ type systems with balanced driving \cite{Baumann2010, baden2014realization}. However, as pointed out by Bhaseen et al. \cite{bhaseen2012dynamics}, relaxation of the balance-driving restriction opens up a much richer dynamics and connects different regions of the phase diagram.  Here we present a first study of the imbalanced driving case.

The general Dicke model is described by the Hamiltonian \cite{bhaseen2012dynamics}, 
\begin{align}
\hat{H} = \omega \hat{a}^\dagger \hat{a} + \omega_0 \hat{J}_z + g_c(\hat{a} \hat{J}_+ + \hat{a}^\dagger \hat{J}_-) + g_p(\hat{a} \hat{J}_- + \hat{a}^\dagger \hat{J}_+ )
\label{eq:general_dicke_model}
\end{align}
where $ \omega$ is the frequency of the field mode, $ \omega_0$ is the frequency splitting between the atomic levels,  $a$ is the annihilation operator for the cavity mode, and $g_c (g_p)$ is co-(counter) rotating atom-light coupling.  The atomic operators $J_\pm , J_z$ are the collective operators for the atomic excitations and satisfy the angular momentum commutation relations  $[J_+, J_-]=2J_z$ and $[J_z, J_\pm]=\pm J_\pm$.  Experimental realizations of this system have utilized a Bose-Einstein condensate (BEC) in an optical cavity \cite{Klinder2015, Baumann2010} or cavity assisted Raman transitions \cite{dimer2007proposed,baden2014realization}.  In the BEC experiments, momentum states of the individual atoms are mapped to the atomic excitations \cite{Klinder2015, Baumann2010}. However, in these experiments the driving strengths $g_c$ and $g_p$ are intrinsically the same.  With the use of cavity assisted Raman transitions, the driving strengths are determined by the intensities of separate laser fields, which are easily tunable. Using this approach, we investigate the imbalanced driving case and demonstrate that this case indeed creates phases very different from the original balanced Dicke model.
\section{Experimental Scheme}
\subsection{Experimental Setup}
\begin{figure}[htbp]
\centering
\fbox{\includegraphics[width=\linewidth]{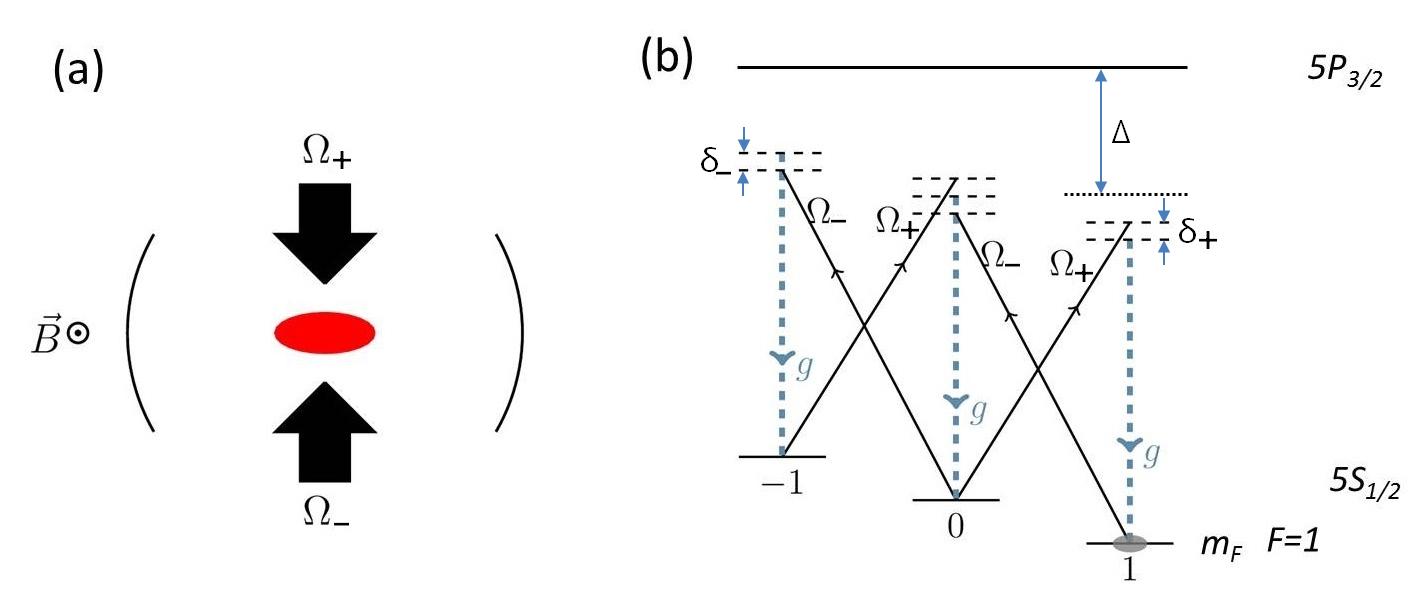}}
\caption{(a) Experimental setup: Atoms in a high finesse cavity are driven from the side by two linearly polarized lasers with the electric fields along the axis of the cavity. A magnetic field is orthogonal to the cavity and the propagation axis of the lasers. (b) Spin-1 Dicke model scheme: each laser field provides a cavity-assisted Raman coupling between $m$ states that can be described by the spin-1 ladder operators, $S_\pm$ as discussed in Sect.~\ref{Theory}.}
   \label{level_scheme} 
\end{figure}

The experimental setup is similar to the one presented in Ref. \cite{arnold2012self} and is illustrated in Fig. \ref{level_scheme}a. An ensemble of $^{87}$Rb atoms are trapped within an optical cavity using an intra-cavity 1560 nm optical lattice, and are driven from the sides with two independently controllable counter-propagating laser beams.
  
The cavity is detuned from the $^2\mathrm{S}_{1/2}$ to $^2\mathrm{P}_{3/2}$ transition by $\Delta=-2\pi\times 127\,\mathrm{GHz}$. Relative to this transition, the cavity parameters are $(g, \kappa, \gamma_a) = 2\pi \times(1.1,0.1, 3)$\,MHz, where  $g$ is the single atom-cavity coupling constant for the $\ket{F=2,m_F = 2}$ to  $\ket{F' =3, m_{F'} = 3}$ cycling transition; $\kappa$ and $\gamma_a$ are half-width-half-maximum linewidth of the cavity and the atomic dipole decay rate, respectively.  A quantization axis is defined by a magnetic field that is orthogonal to the optical axis of the cavity and gives a measured Zeeman splitting between neighboring magnetic sub-levels of $\omega_z=2\pi\times 1.577\,\mathrm{MHz}$ corresponding to a magnetic field of $\approx 0.225\,\mathrm{mT}$.  The lasers are linearly polarized with the electric field along the axis of the cavity and propagate orthogonal to the magnetic field.  In this configuration, differential stark shifts and differential dispersive shifts between different $m$ states are negligibly small.  The laser polarizations are described by an equal superposition of circular polarizations $\sigma^\pm$.  When the laser fields are tuned close to a cavity-assisted Raman resonance, as illustrated in Fig.~\ref{level_scheme}(b), only one component contributes significantly to the dynamics: one beam providing a coupling through the $\sigma^+$ component, and the other through the $\sigma^-$ component.  Hence, in what follows, we denote the parameters associated with each laser by the subscript $\pm$.
\subsection{Theoretical Model}
\label{Theory}
Derivation of the Hamiltonian for the level scheme in Fig.~\ref{level_scheme}(b) is similar to that given in \cite{dimer2007proposed} and we summarize the relevant results.  Since the detuning of the lasers from the excited states is large, we can adiabatically eliminate the $^2\mathrm{P}_{3/2}$ levels and neglect the hyperfine structure.  Neglecting far off-resonant cavity-assisted Raman transitions, the system, in a suitable rotating frame, can then be described by the master equation (taking $\hbar=1$)
\begin{equation}
\dot{\rho} =-i[H, \rho] + \kappa(2a\rho a^\dagger-a^\dagger a\rho-\rho a^\dagger a)
\label{eq:master_equation} 
\end{equation}
with
\begin{equation}
H =  \omega a^\dagger a + \omega_0 \hat{J}_z + \frac{\lambda_-}{\sqrt{2N}}(a J_+ + a^\dagger J_-) + \frac{\lambda_+}{\sqrt{2N}}(a J_- + a^\dagger J_+).
\label{eq:hamiltonian} 
\end{equation}
The collective operators $J_\pm, J_z$ are the usual angular momentum operators given by,
\begin{align}
J_\pm&=\sum_{j=1}^{N} S_{\pm,j} =\sum_{j=1}^{N}\sqrt{2}\big(\ket{\pm 1_j}\bra{0_j}+\ket{0_j}\bra{\mp 1_j}\big),\\
J_z&=\sum_{j=1}^{N} S_{z,j} = \sum_{j=1}^{N}\big(\ket{1_j}\bra{1_j} -\ket{-1_j}\bra{-1_j}\big), 
\end{align}
where $S_{j,\pm},S_{j,z}$ are the spin 1 angular momentum operators for each atom. The model parameters $\omega_0, \omega,$ and $\lambda_\pm$ are given by
\begin{equation}
\omega=\left(\omega_c -\frac{\omega_++\omega_-}{2}+\omega_d\right), \quad \omega_0=-\left(\omega_z +\frac{\omega_+-\omega_-}{2}\right), \quad \lambda_\pm =-\frac{\sqrt{2N}g\Omega_\pm}{24\Delta},
\end{equation}
where 
\begin{equation}
\omega_d = \frac{2}{3}\frac{Ng^2}{\Delta},\quad \mathrm{and} \quad \Omega_\pm=\frac{E_\pm\bra{{^2\mathrm{P}_{3/2}},3,3}\mathbf{r}\cdot\hat{\sigma}_+\ket{{^2\mathrm{S}_{1/2}},2,2}}{\hbar}.
\end{equation}
In these expressions, $E_\pm$ are the electric field amplitudes of each laser, and $\omega_\pm$ the associated optical frequency.  Note that the dispersively corrected detuning, $\delta_\pm$, of each laser from the cavity-assisted Raman resonance between the $m$ and $m\pm1$ states is given by
\begin{equation}
\delta_\pm=\omega_\pm\pm \omega_z-(\omega_c+\omega_d)=-(\omega\pm\omega_0).
\label{eq:the laser beams}
\end{equation}
Omitted terms associated with couplings from the $\sigma^\mp$ component of the $\Omega^\pm$ beam are detuned by $\delta_\pm\mp2\omega_z$, so eq.~\ref{eq:hamiltonian} is only valid provided $|\delta_\pm|\ll 2\omega_z$.  We also note that the amount of spontaneous emission arising from each laser is completely determined by $\lambda_\pm$ and is given by
\begin{equation}
\gamma_s=\frac{96\lambda_\pm^2}{NC\kappa},
\label{spontaneousEmission}
\end{equation}
where $C=g^2/\kappa\gamma_a$ is the single atom cooperativity.  

It is worth noting that the description here is equally valid for the $F=2$ case.  In this case the individual atomic excitations are given by spin 2 operators and the only changes to the final Hamiltonian are: a change of sign to $\lambda_\pm$ due to a change in sign of relevant matrix elements, a change of sign of $\omega_z$ due to a change in sign of the Lande g-factor, $g_F$, and the replacement $N\rightarrow 2N$ in the expressions for $\lambda_\pm$ and $H$ given in Eq.~\ref{eq:hamiltonian} due to the increased spin.  Higher spin models have also be considered in \cite{spin1Scott}.

\section{Experimental Implementation}
An experiment starts by preparing a set number of atoms in $\ket{F=1, m_F=1}$ with a well-defined temperature and this is achieved as follows.  First a magneto-optical trap (MOT) is formed 15\,mm above the cavity.  The atoms are then pumped into the $F = 1$ hyperfine manifold and transferred to a single-beam 1064\,nm dipole trap that is overlapped with the MOT.  Typically, around $5\times10^6$ atoms are loaded into the dipole trap.  Using a motorized translation stage, the beam is then moved down 15\,mm over one second to bring the atoms into the cavity. In the cavity, the atoms are optically pumped into $\ket{F=1, m_F=1}$ with approximately 94(2)\% efficiency. The power of the 1064\,nm beam is then adiabatically lowered in 350\,ms to transfer the atoms into the 1560\,nm intra-cavity optical lattice, which is 219(4) $\mu$K deep. The number of atoms transferred to the intracavity optical lattice is determined non-destructively by measuring the dispersive shift, $\omega_d$, of the cavity with an accuracy of $\sim5\,\mathrm{kHz}$. 

A fixed atom number of $2.0(1)\times 10^5$ is maintained run-to-run using a field programmable gate array (FPGA) which triggers the experiment once a set value is reached.  Explicitly, the cavity probe beam is set to a fixed frequency slightly less than the maximum dispersive shift. We then monitor the cavity transmission using a single photon counting module (SPCM) and record the count with the FPGA.  As the atoms are lost due to background collisions, the cavity is moved into resonance with the probe beam, increasing the output photon count rate. When the count rate reaches a preset threshold, the FPGA is triggered and clocks out the rest of the time sequence.  This procedure provides a high degree of repeatability in the experiment with evaporation ensuring a well defined temperature relative to the depth of the intracavity 1560\,nm optical lattice, and triggering off a set dispersive shift fixing a well defined atom number.  Remaining variation in the dispersive shift can be further reduced by post-selection, as it is also measured \emph{in situ}.

In the experiment, all lasers are referenced to a high finesse transfer cavity with a linewidth of $\sim 50\,\mathrm{kHz}$ at both $780\,\mathrm{nm}$ and $1560\,\mathrm{nm}$.  In addition, the experiment cavity is locked to the $1560\,\mathrm{nm}$.  This allows all laser detunings to be accurately set relative to the empty cavity resonance at either $1560\,\mathrm{nm}$ or $780\,\mathrm{nm}$.  Complete specification of the model parameters then requires a measurement of $\omega_z$ and a characterization of $\lambda_\pm$ in addition to the \emph{in situ} measurement of the dispersive shift.  This is done in a separate single beam experiment.

When $\lambda_- = 0$, the Hamiltonian reduces to a Tavis-Cummings interaction.  Weak probing of the cavity then provides an avoided crossing with a splitting that is determined by $\lambda_+$.  Fitting the cavity transmission as a function of both the probe detuning with respect to the cavity and the cavity detuning with respect to the Raman resonance then allows us to extract both the coupling strength $\lambda_+$ and the splitting $\omega_z$.  At a beam power of $1\,\mathrm{mW}$ and $\omega_d=2\pi\times 0.85\,\mathrm{MHz}$ we obtain $\lambda_+=2\pi\times50.3(1.2)\,\mathrm{kHz}$ and $\omega_z=2\pi\times 1.577(2)\,\mathrm{MHz}$.  The value of $\lambda_+$ can then be scaled to other powers or atom numbers (dispersive shifts).  Since the two coupling lasers are mode matched via fiber to fiber coupling, $\lambda_+=\lambda_-$ for equal laser powers.

With the parameters of the model fully characterized, we are able to explore the phase diagram. We run the experiment for various ratios of the coupling strengths $\lambda_+/\lambda_-$ with $\omega$ and $\omega_0$ kept at 100(5) kHz and -77(2) kHz, respectively\footnote{Strictly, $\omega_0<0$ as we prepare the atoms in $\ket{+1}$.  The more conventional sign choice amounts to an inconsequential coordinate change.}. For each run, after 10 ms of the FPGA trigger, the laser fields are switched on at low power and then ramped over $200\,\mathrm{\mu s}$ with a fixed ratio of beam powers consistent with that set by ratio, $\lambda_+/\lambda_-$.  This ensures beam powers are properly stabilized once the cavity dynamics takes hold. The beams remain on for 3\,ms and the output from the cavity is collected using an SPCM. 

\section{Results and Discussions}
\begin{figure}[htbp] 
 \begin{center}
 \includegraphics[width= \textwidth]{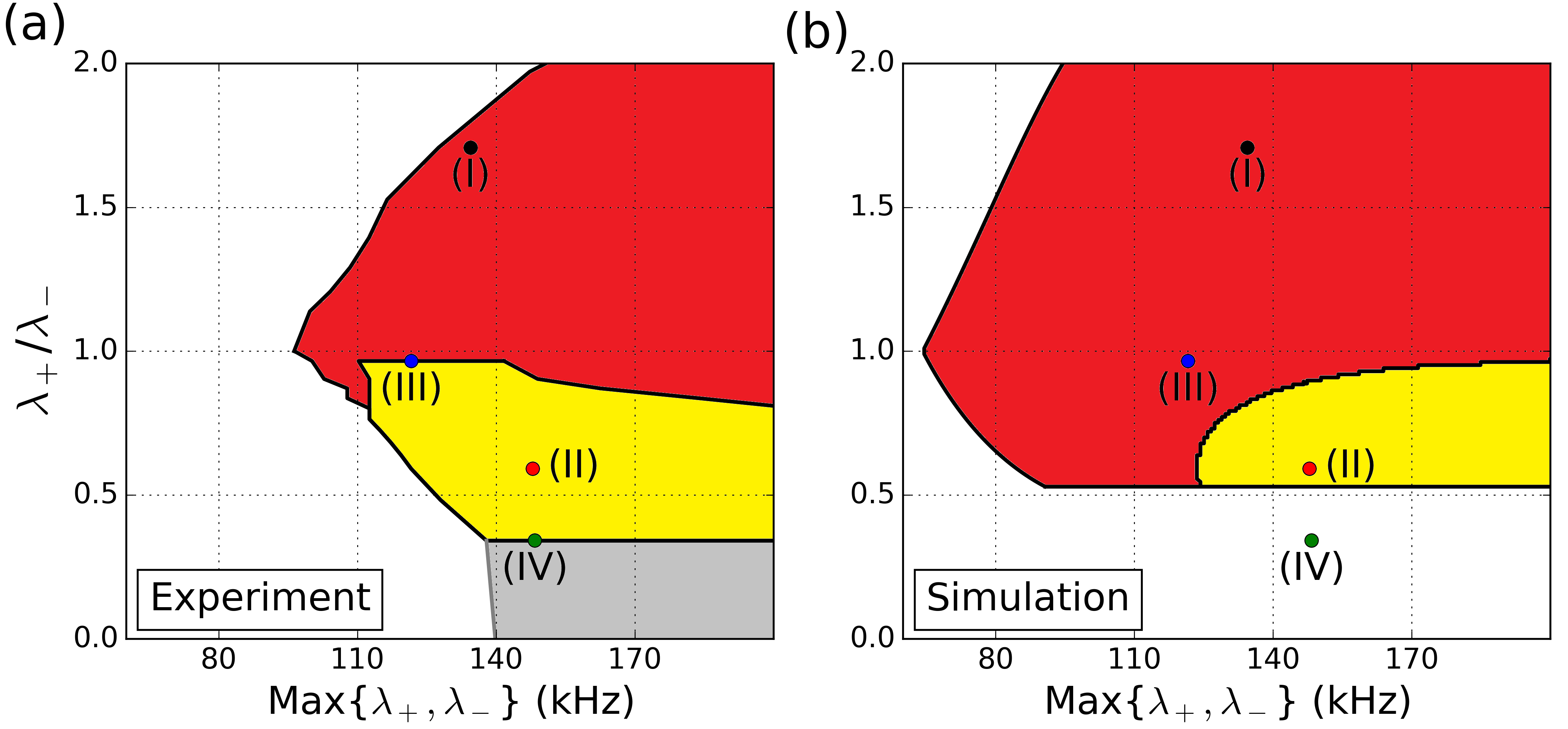}
 \includegraphics[width=  \textwidth]{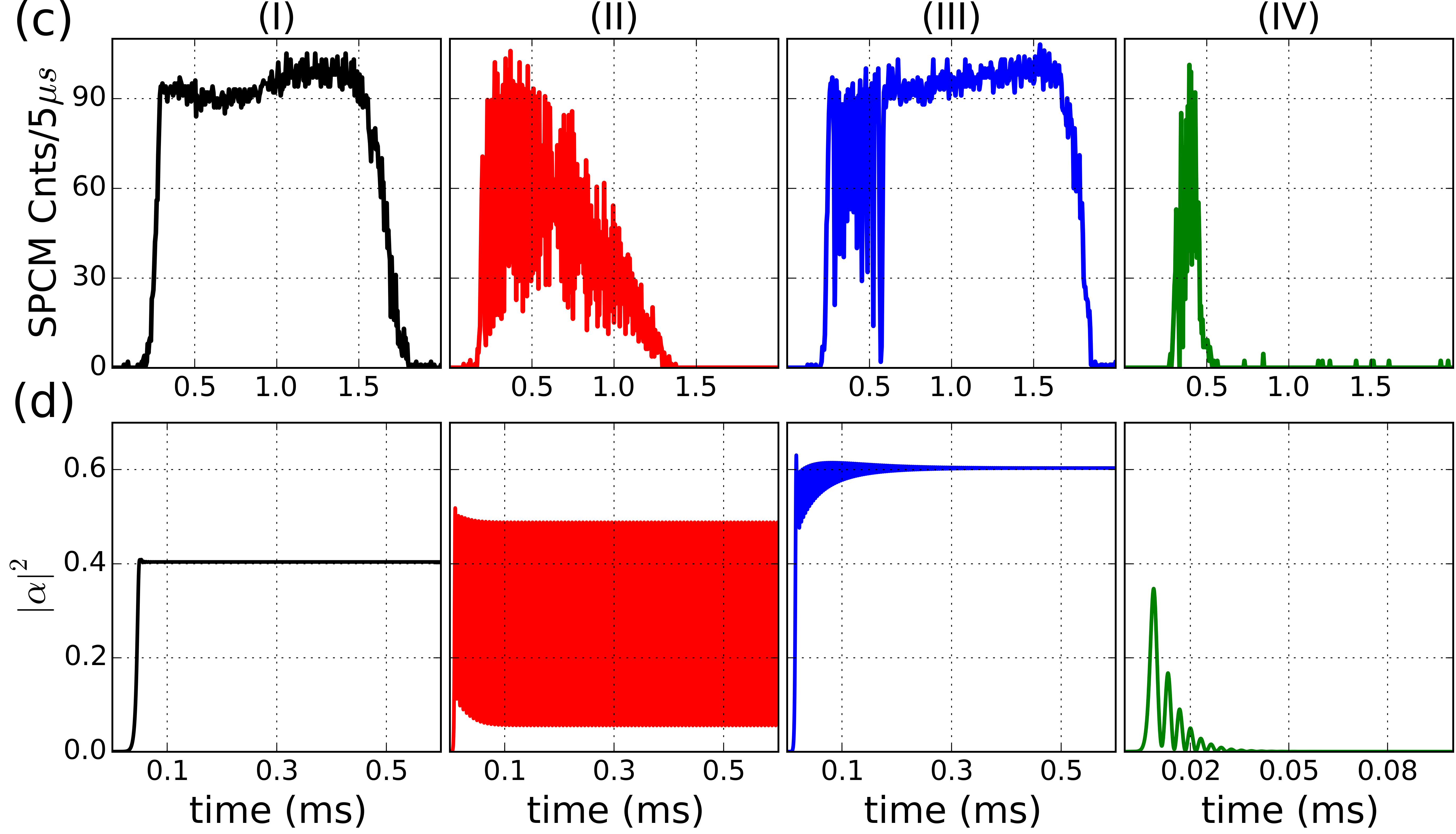}
 \caption{Experimental (a) and theoretical (b) phase transition diagrams for the spin-1 Dicke system. The vertical axis of (a) and (b) represents the ratio of the coupling strength ($\lambda_+/\lambda_-$), and the horizontal axis represents the higher value of the coupling strength $\mathrm{max}\{\lambda_+,\lambda_-\}$. The white, red and yellow regions represent the trivial, normal super-radiant and oscillatory states, respectively.  Designation of states for the experimental results are based on the cavity output and its  Fourier transforms as discussed in the text. Theoretical results are based on the steady state of the system after a long integration time.   In (c) and (d), representative cavity outputs (I) to (IV) are given for the corresponding locations indicated in (a) and (b).}
\label{fig:main}
\end{center}
\end{figure}

Outputs from the cavity for a selection of coupling strengths is given in Fig.~\ref{fig:main}(c) and a complete phase map is shown in Fig.~\ref{fig:main}(a).  The white, red and yellow regions represent the trivial state, the normal super-radiant state and the oscillatory super-radiant state, respectively.  The trivial state corresponds to zero output from the cavity with all atoms in either $\ket{+1}$ or $\ket{-1}$.  The super-radiant state is characterized by a $1\sim2\,\mathrm{ms}$ output pulse with a steady amplitude as in the output indicated in (I).  The oscillatory states are characterized by an obvious oscillation in the cavity output as in (II), which is identified by a strong peak in the Fourier transform of the cavity output signal.  The grey region corresponds to short pulse lengths of around $200\,\mu\mathrm{s}$, as in (IV), which we interpret as a cavity-assisted Raman transfer from a single beam moving the spin population to a trivial steady state.  It should be noted that the SPCM saturates at 100 counts/5 $\mu$s which does not allow us to compare the amplitude of the signal.  Also the gate time of the SPCM is 5 $\mu$s which limits the interpretation of the oscillation frequencies for the oscillatory states in the experiment. 

 
In Fig~\ref{fig:main}, we also provide a comparison with theory, which we obtain from the semi-classical equations of motion,
\begin{subequations}
\begin{align}
\dot{\alpha}&= -\kappa \alpha - i\omega \alpha - i\lambda_- \beta-i\lambda_+ \beta^*,\\
\dot{\beta} &= - i\omega_0 \beta + 2i\lambda_- \alpha w+2i\lambda_+ \alpha^* w,\\
\dot{w} &= i \lambda_-(\alpha^* \beta - \alpha \beta^*) +i \lambda_+(\alpha \beta - \alpha^* \beta^*),
\end{align}
\label{eq:semiclassical} 
\end{subequations}
\text{where, } 
\begin{equation} 
\alpha = \frac{\langle a \rangle}{\sqrt{2N}}, \quad \beta = \frac{\langle J_-\rangle}{2N}, \quad w = \frac{\langle J_z\rangle}{2N}.
\label{eq:beta-gamma_relation} 
\end{equation}
These follow from the master equation (\ref{eq:master_equation}), with Hamiltonian (\ref{eq:hamiltonian}), by neglecting quantum fluctuations and imposing the factorizations
\begin{equation}
\langle (a+a^\dagger) J_z\rangle  \rightarrow \langle (a+a^\dagger)\rangle \langle J_z\rangle,\quad \langle (a+a^\dagger) J_\pm\rangle \rightarrow \langle (a+a^\dagger)\rangle \langle J_\pm\rangle.
\end{equation} 
Initial conditions used for simulations are slightly perturbed from $(\alpha, \beta, w) = (0, 0, 0.5)$ and the equations are integrated until a steady state or stable limit cycle can be identified.  We note that some of the boundaries between the trivial and super-radiant phases can be determined analytically from Eqs~\ref{eq:semiclassical} \cite{bhaseen2012dynamics}).The associated phase diagram is shown in Fig.~\ref{fig:main}(b) and the integrated cavity output for a selection of  coupling strengths in Fig.~\ref{fig:main}(d).  These can be compared with the corresponding experimental results in Fig.~\ref{fig:main}(a) and (c) to which they show reasonable qualitative agreement.

Differences between experiment and theory are apparent in the finite duration of the cavity output as seen in the experiments, and in the location of the boundaries in the phase space diagrams.  The finite duration of the cavity output is likely due to dephasing effects such as spontaneous emission and collisions, which results in a decay of the collective spin \cite{torre2016dicke,gelhausen2016many,bohnet2012relaxation}.    From Eq.~(\ref{spontaneousEmission}), spontaneous emission rates are on the order of $100\,\mathrm{s}^{-1}$ per beam, and we estimate a collision rate of $800\,\mathrm{s}^{-1}$ in the intracavity $1560\,\mathrm{nm}$ lattice. These dephasing rates are consistent with the millisecond timescales observed in the experiment.  

The finite duration of the pulse and the manner in which we infer an oscillatory state influences the position of the boundary between oscillatory and normal super-radiant states, particularly at lower values of the coupling strength.  In this region, solutions show an initial oscillation that is sustained over a fraction of a millisecond or more but eventually decays to a steady state as evidenced by cavity outputs (III).  Experimentally these are identified as an oscillatory state in that they exhibit a clear peak in the Fourier spectrum.  At higher coupling strengths, simulations near to the boundary show oscillatory outputs that have small amplitude oscillations upon a much larger DC background.  Experimentally, these oscillatory states are seen as a steady state as the oscillations are masked by either noise or saturation of the SPCM.

The lower boundary at $\lambda_+/\lambda_-=0.529$ is also difficult to determine.  Below the boundary simulations show a short pulse output that corresponds to the atoms being transferred from $J_z=N$ to $J_z=-N$.  Experimentally, this boundary is much less sharp with pulses increasing in duration as one approaches the boundary from below.  These shorter pulses can also show a marked peak in the Fourier spectrum which compromises the ability to determine the boundary.  This is illustrated in the cavity outputs (IV) in Fig.~\ref{fig:main} (c) and (d).

The boundary to the left is also significantly shifted to the right indicating that a higher value of the coupling is needed for a phase transition to occur.  Decoherence mechanisms have been explored to explain this effect \cite{torre2016dicke,gelhausen2016many, bohnet2012relaxation}.  However decoherence rates needed to explain the larger threshold are substantially larger than the spontaneous emission and collision rates present in the experiment.  It is not obvious why this might be,  but we do note that our theoretical model does not account for the thermal distribution of atoms in the cavity. Experimentally, determination of the dispersive shift and calibration of the coupling strengths $\lambda_\pm$ inherently take into account the atomic spatial distribution.  However, it is not clear that this adequately accounts for this effect in the Dicke-model implementation.  Doppler shifts also have a significant effect on the value of $\omega_0$.  Regardless of these limitations, our results show reasonable qualitative agreement with the simple Dicke model given by Eq.~(\ref{eq:hamiltonian}) and demonstrates the predictions made in \cite{bhaseen2012dynamics}.

\section{Conclusion}
In this paper, we have presented a realization of a spin-1 Dicke model using cavity-assisted Raman transitions between Zeeman sub-levels of an $F=1$ hyperfine state. Our realization provides wide tunability of the Dicke model parameters and has allowed us to explore the more general imbalanced driving condition as explored in \cite{bhaseen2012dynamics}. We have prepared the phase transition map for the coupling strength ratio of the Raman beams ($\lambda_+/\lambda_-$) from 0 to 2 which shows qualitative agreement with the ideal Dicke model.   

Our implementation also provides an experimentally convenient realization of the Dicke-model.  It eliminates an additional term in the Hamiltonian associated with a differential dispersive shift of the cavity between differing atomic states.  It also eliminates differential AC stark shifts from the coupling beams which simplifies the characterization of the model parameters.  Furthermore the scheme presented in this work could be easily extended to include more atomic states and realize higher spin Dicke type models. 
\section*{Acknowledgements}
This research is supported by the National Research Foundation, Prime Ministers Office, Singapore and the Ministry of Education, Singapore under the Research Centres of Excellence programme (partly under grant No. NRF-CRP12-2013-03).  S. J. Masson and A. S. Parkins acknowledge support from the Marsden Fund of the Royal Society of New Zealand (Contract No. UOA1328).  They also acknowledge the contribution of NeSI high-performance computing facilities to the results of this research. NZ's national facilities are provided by the NZ eScience Infrastructure and funded jointly by NeSI's collaborator institutions and through the Ministry of Business, Innovation \& Employment's Research Infrastructure programme.



\bibliographystyle{osajnl}
\bibliography{bib.bib}







\end{document}